\documentclass{iau}
\usepackage{graphicx,natbib,amsmath,color}
 
\newcommand{\atlas}{{\sc Atlas$^\mathrm{3D}$}}

\newcommand{\hi}{{\sc H\,i}}
\newcommand{\solmass}{$\rm M_{\odot}$}
\newcommand{\msunsqpc}{$\rm M_{\odot} pc^{-2}$}
\newcommand{\error}{$\pm$}
\newcommand{\htoo}{H$_2$}
\newcommand{\mhtoo}{M(H$_2$)}

\newcommand{\mhtoomstar}{M(H$_2$)/M$_\star$}

\newcommand{\mhi}{M({\sc H\,i})}
\newcommand{\mstar}{M$_\star$}
\newcommand{\asec}{$''$}
\newcommand{\arcsec}{$''$}
\newcommand{\arcmin}{$'$}
\newcommand{\kms}{km~s$^{-1}$}

\title{The Recent Evolution of Early-Type Galaxies as Seen in their Cold Gas}

\author[Young]{Lisa M. Young$^1$}

\affiliation{$^1$Physics Department, New Mexico Tech, 801 Leroy Place, Socorro NM,
87801, USA \\
email: {\tt lyoung@physics.nmt.edu} }

\pubyear{2014}
\volume{309}
\jname{Galaxies in 3D across the Universe}
\editors{B. L. Ziegler, F. Combes, H. Dannerbauer, M. Verdugo, eds.}

\begin{document}

\maketitle

\begin{abstract}
I present an overview of new observations of atomic and molecular gas in 
early-type galaxies, focusing on the \atlas\ project. 
Our data on stellar kinematics, age and metallicity, and ionized gas
kinematics allow us to place the cold gas into the broader context of early-type galaxy
assembly and star formation history.  The cold gas data also provide valuable
constraints for numerical simulations of early-type galaxies.

\keywords{galaxies: elliptical and lenticular, cD - galaxies: evolution - galaxies: formation}
\end{abstract}

\firstsection
\section{Introduction}

Two of the major interrelated challenges of galaxy formation are 
understanding the basic origin of
the Hubble sequence (spiral vs.\ elliptical and lenticular galaxies) and the
present-day dichotomy in galaxy colors (the red sequence vs.\ the blue cloud or 
star-forming sequence).
Most early-type (elliptical and lenticular) galaxies are poor
in cold gas; yet their stars did form, albeit a long time ago, in a precursor that
was probably gas-rich and blue.  In this context, then, the story of what happened
to the cold gas in those precursors is a crucial part of the evolution of early-type
galaxies.  The current gas contents of early-type galaxies can offer valuable
quantitative benchmarks for numerical simulations of galaxy evolution.

Early-type galaxies probably moved onto the red sequence as they lost their cold gas
to consumption (star formation),  AGN and star formation feedback, and environmental
processing in groups and clusters.  On the other hand, they could also have acquired 
cold gas through cooling out of a hot medium and through mergers.  
Their cold gas supports some current star formation activity, and the rates
and efficiencies of that process (in comparison to the behaviors of spirals) can
offer some useful constraints on star formation in general. 
These are some of the issues addressed through observations of nearby early-type
galaxies in the \atlas\ project.

\section{Sample and Observations}

The \atlas\ sample is a complete volume-limited sample of early-type galaxies
nearer than 42 Mpc and brighter than $M_K = -21.5,$ i.e.\ 
having stellar masses greater than $M_\star = 10^{9.9}$
\solmass\ \citep{paper1}.   
The early-type galaxy sample is actually drawn from a parent sample which has no
color or morphological selection, and
the 260 galaxies lacking spiral structure form the basis of the project.
The core of the observations consists of 
integral-field optical spectroscopy over a field of at least 33\asec $\times$
41\asec.  For these nearby galaxies, 80\% of the sample has spectroscopic
coverage out to $0.8 R_e$ or farther \citep{paper3}.
While the wavelength coverage of the spectra is not large by current standards, the
data do offer stellar kinematics, [O III], H$\beta$ and occasionally other emission lines, and line indices
sensitive to age, metallicity, and $\alpha$ enhancement.
This is all useful for inferring both the star formation and the assembly histories of 
the galaxies, in addition to the ionized gas excitation and kinematics.

\atlas\ galaxies primarily inhabit the red sequence, as expected, though a small 
percentage are blue, especially at masses $M_\star < 10^{10.5}$ \solmass\ 
\citep{paper1,CMDpaper}.
A discussion of their morphological characteristics is found in \citet{paper7}.
Basically, the majority of early-type galaxies are disky, fast rotating, oblate
galaxies, 
and relatively few, primarily at high masses, are spherical or triaxial and slowly 
rotating \citep{paper2,paper3,paper24}.
The two-dimensional kinematic data highlight the incidence of multiple kinematic
subcomponents, decoupled or even counterrotating cores, and kinematic twists within the
galaxies \citep[see especially][]{paper2}.
Ongoing work also involves deep optical imaging \citep{paper9} to give better
limits on recent interactions.

An initial CO survey with IRAM 30m telescope probed the molecular content in the 
central ~20\arcsec\ of the \atlas\ galaxies, giving detection limits in the range
$10^7$ to $10^8$ \solmass\ \citep{paper4}.
Some more exotic high density molecular tracers have also been observed in a few
cases \citep{bayet12,crocker-HD,bayet13,davis-cs}.
\citet{carma} present interferometric CO observations of most of the CO detections,
at a typical spatial resolution of 5\arcsec\ (300 pc to 1 kpc) over a 1\arcmin\ (5 -- 11 kpc) 
field of view.
\citet{paolo} also present interferometric \hi\ observations of 170 of the \atlas\
galaxies from the Westerbork array; those data cover a 30\arcmin\ field and a
velocity range of 4000 \kms\ at typical resolutions of 35\arcsec, and they have the
sensitivity to detect \hi\ column densities of a few $10^{19}$ cm$^{-2}$.
For a modest-sized cloud of gas this corresponds to mass limits of $10^6$ to $10^8$
\solmass.

\section{Abundant Cold Gas in Early-Type Galaxies}

From the observations noted above, the detection rate of molecular gas in massive 
early-type galaxies is 22\% \error 3\% \citep{paper4}.  To place this value in
context it is important to remember that the sample selection is not biased by
far-IR detections or even  by selection in the $B$ band, as were many previous 
surveys in early-type galaxies.
The \htoo\ masses fall in the range $10^{7.1}$ to $10^{9.3}$ \solmass, or
log \mhtoomstar\ in the range $-3.5$ to $-1.1$.  For the galaxies with CO maps,
the molecular gas is usually concentrated in the central kpc or so \citep{davis13}
in disks, rings, bars, and irregular configurations \citep{carma}.
The detection rate of \hi\ is 40\% outside of the Virgo Cluster; \hi\ masses range
from $10^{6.3}$ \solmass\ to $10^{10}$ \solmass, and the atomic gas is found in isolated clouds,
regular disks (sometimes many tens of kpc in diameter) and disturbed configurations
\citep{paolo}.

The \atlas\ project thus has
the largest set of cold gas maps ever assembled for early-type galaxies.
Atomic and/or molecular gas in quantities of 
$10^7$ to $10^9$ \solmass\ is surprisingly
common in these galaxies, being found in about 50\% of them.  Often the gas-rich
galaxies are detected in both phases, but not always; and 
the galaxies have values of \mhtoo/\mhi\ that range from $10^{-2}$ to $10^2$, even considering just
the ones detected in both phases.

\section{Which kinds of ETGs have cold gas?}

\begin{figure}
\centering
\includegraphics[scale=0.65]{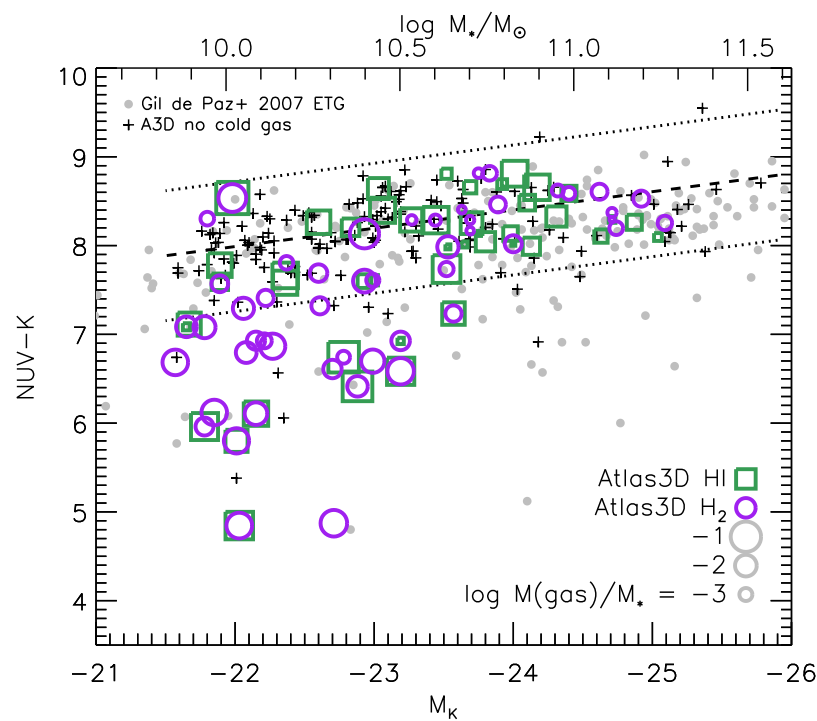} 
\caption{NUV-K color-magnitude diagram for the \atlas\ sample.  The dashed line is
the red sequence ridge line and the dotted lines are $2\sigma$ redder and bluer, as
described in \citet{CMDpaper}. \hi\ and \htoo\ masses are indicated by the symbol
sizes. }
\label{fig:fig1}
\end{figure}

Figure \ref{fig:fig1} shows a color-magnitude diagram of the \atlas\ 
galaxies, and it demonstrates that molecular gas detection rates are still strong
among red sequence galaxies, particularly at high stellar masses where the star
formation rates are not large enough to move the galaxies off the red sequence in
integrated colors.
\citet{CMDpaper} have also discussed the reddening effects of dust, which are modest.
Careful statistical analysis shows that the molecular mass detection rate is not a
function of stellar
mass over the observed ranges, nor is the \htoo\ mass distribution function
\citep{paper4}.

While the molecular masses of early-type galaxies do not correlate with stellar
mass, they do correlate with structural parameters.
\citet{paper20} show that, at a fixed stellar mass, the early-type galaxies with
low stellar velocity dispersion and large effective radius (i.e.\ with more dominant
disky stellar components rather than spheroids) have larger molecular masses.
It is tempting to assume that this might be because the molecular gas disk gives
birth to a young, disky stellar component.  And while that is partly true, 
the currently observed
molecular masses are usually only 1\% of the stellar masses, or less, so the
young stellar disks are probably too small to drive the whole-galaxy bulge/disk
ratios.  They must be driven by older stellar disks.  In addition, as we will discuss below, much of the molecular gas is
kinematically misaligned with respect to the stars.

There are some interesting differences in the atomic and molecular contents of
early-type galaxies.
\citet{paper2} 
show that 
about 86\% of early-type galaxies are the oblate fast rotators.  
These galaxies can grow through gas-rich minor mergers and associated {\it in situ}
star formation, or through recent major mergers with the appropriate geometry to
preserve a large net angular momentum \citep{naab}. 
They are detected at about the same rate in both CO and \hi\ \citep{paolo}.
More rare are the slow rotators, which 
can be formed from recent mergers with special geometries ensuring near-cancellation
of the net angular momentum, and the spherical non-rotators, which are proposed to
grow through a large number of isotropically oriented minor mergers \citep{naab}.
Slow- and non-rotators are rarely detected in CO but are detected in \hi\ at about
the same rate as the fast rotators are detected in \hi\ \citep{paolo}. 
This information suggests that at least in some cases the origin of the atomic gas
is different from that of the molecular gas, or perhaps the galaxy structure
affects the formation or longevity of the molecular phase.  \citet{sarzi13}
also discusses some effects of the galaxy structure on the hot gas content of the \atlas\
galaxies.

We conclude that the quenching of a galaxy and the approach to the red sequence do 
not always involve the loss of all, or even most, of the cold gas from an early-type
galaxy.  Alternatively, perhaps some cold gas is reacquired during or after the 
approach to the red sequence.  The presence of molecular gas in an early-type
galaxy is strongly connected to the presence of a dynamically cold, disky stellar
component, and the connection may be partly (but not entirely) due to the 
formation of a young stellar disk out of said molecular gas.
Contrary to this, atomic gas is more indiscriminate, being found in 
approximately equal quantities in both
the disky fast rotators and the bulge-dominated slow rotators, and often not associated
with any notable star formation.

\section{Angular Momenta as clues to the history of the gas}

Early-type galaxies have much more interesting and diverse stellar kinematics in
their inner regions than spirals do --- as is abundantly clear from the examples
shown in \citet{paper2}.
The comparisons between gas and stellar kinematics are also important clues to the
evolution of these galaxies.
For example, \citet{davis11} plot the misalignment angle between stellar and
ionized or molecular gas kinematics; the distribution peaks at zero but has a long flat tail
all the way to $180^{\circ}$ (retrograde gas).
Stellar--gas kinematic misalignments are common in molecular and ionized gas, 
occurring in at least 35\% of early-type fast
rotators \citep[see also][]{sarzi10}.  
Similarly, half of all \hi-detected fast rotators
have that \hi\ kinematically misaligned with respect to the stars \citep{paolo-kin}.
These papers argue that gas at misalignments $> 30^{\circ}$ cannot be simply recycled
material from internal stellar mass loss, because it has the wrong kinematics.
Perhaps it could be brought in to the galaxy in an accreted satellite, or perhaps
(as discussed below) it cooled out of hot gas at large radii in the galaxy halo,
where the hot gas is vulnerable to many external torques.
If these arguments apply to all of the misaligned gas plus some portion of the aligned
gas as well, they apply to something like $\gtrsim 50\%$ of the cold gas in nearby
early-type galaxies.

There is also some evidence for external gas accretion from the metallicity
measurements.
The \atlas\ sample galaxies follow a well-established
stellar mass--metallicity relation, with a few notable outliers at markedly low
metallicity (but enhanced $\alpha$-element abundances) for their mass.  
These also tend to be blue and rich in
molecular gas.  \citet{richard-sf} postulate that these are cases in which low
metallicity gas was accreted, so the population of younger stars with lower metallicity
is biasing the ``average" metallicity estimate.

\section{Comparison to simulations}

Past work on simulating the cold gas content of galaxies has mainly focused on the
gas-rich spiral galaxies \citep[e.g.][]{lagos-11}.  The simulations of early-type 
galaxies are more challenging if for no other reason than the quantities of 
cold gas are generally smaller, requiring higher fidelity.
\citet{paolo-kin} made high resolution re-simulations 
of dark matter halos extracted from a larger
$\lambda$CDM simulation, implementing gas cooling, star formation and supernova 
feedback. 
Cooling is not carried out to molecular gas temperatures, but \hi\ is converted to
\htoo\ by fiat at surface densities of 10 \msunsqpc.
Not surprisingly, the 
details of the feedback prescription 
have strong influence on the final gas distribution
and kinematics, and even in some cases on the stellar kinematics.  
There is significant tension between these simulations and the observations in the
sense that the simulations produce too much prograde cold gas and not enough misaligned
cold gas.

\citet{lagos-14} describe a semi-analytical model which follows the 
quantities of cold gas that cooled out of the hot halo,
were brought in by mergers, returned to the ISM from internal stellar mass loss, and
consumed by star formation.
They stress the co-evolution of the hot and cold gas, emphasizing that an improved
treatment of gentle ram pressure stripping of the hot halo is necessary to reproduce
the observed cold gas contents of early-type galaxies.  
A work in preparation also addresses the kinematic misalignments of the cold gas in
early-type galaxies; there are not enough minor mergers to explain all of the
misaligned gas, but cooling out of a hot halo might contribute misaligned cold gas if
the hot halo has been torqued so that it is also misaligned with the stellar body of
the galaxy (Lagos 2014, private communication). 

\section{Star formation and galaxy dynamics}

For some years it has been suspected that the empirical star formation ``laws"
(predicting the star formation rate produced by a given quantity of cold gas) 
might be somewhat different in early-type galaxies than in spirals, due to the
systematically different gravitational potentials
\citep[e.g.][]{kawata,martig09}. 
Indeed, hydrodynamical simulations of \citet{martig13} found that the molecular gas in
spherical galaxies should not reach the same high local densities as the molecular gas
in spirals, and they predicted something like factors of 2 lower star formation efficiencies
in early-type galaxies.  These kinds of factors of two could prove to be important as
simulations of galaxy evolution move into precision mode.

Interestingly, \citet{davis-sf} observe that the star formation 
efficiency of early-type galaxies
correlates with a measurement of whether most of the gas is in the rising or the flat
part of the rotation curve.  Specifically, where most of the molecular gas is in the
rising part, the efficiency is lower by factors of a few.  
These types of observations offer the opportunity to make empirical tests of the star
formation ``laws" beyond the spiral galaxies where they were developed, so they should help
refine the prescriptions used in simulations.

\section{Summary}

There is more cold gas in early-type galaxies than most of us expected.  Approximately
50\% of massive early-type galaxies (\mstar\ $\gtrsim 10^{10}$ \solmass) contain 
$10^7$ to $10^9$ \solmass\ of \hi\ and/or \htoo.  The atomic and molecular phases are
not always found together; for example, Virgo Cluster members are more likely to be
detected in \htoo\ but not in \hi.

Because many of these cold-gas-rich early-type galaxies are found on the red sequence,
and the red colors are {\it not} solely the effect of dust associated with the cold gas, we infer that
`red sequence' and `cold gas' are not mutually exclusive properties; the approach to the red
sequence does not always involve the loss of all cold gas.  Alternatively, some red
sequence galaxies may reacquire gas after going onto the red sequence.

Much of the cold gas in early-type galaxies is kinematically misaligned with respect to
the stars.  Because of the misalignment it is apparently not a simple recycling of gas from stellar mass loss
directly back into a cold gas phase.
It remains to be seen --- and would be fascinating to know --- how the cold
gas is aligned (or not) with respect to the hot gas, as it becomes increasingly obvious
that hot and cold phases probably do interact with each other.
We have posited that a significant portion of the cold gas in early-type galaxies is
accreted either with incoming satellite galaxies in minor mergers, or has cooled from
the intergalactic medium or a hot halo.  

The cold gas contents and kinematics of early-type galaxies are thus shown to be
sensitive probes of their evolution.
The diversity in the stellar kinematics, gas contents, and gas kinematics of early-type
galaxies emphasizes the diversity in their evolutionary paths.
Even the diversity in the star formation properties of early-type galaxies can help
improve our understanding of the star formation process.
These features are now beginning to give useful constraints on numerical simulations.

Additional work on the hot gas in early-type galaxies is needed, especially regarding
the thermal and dynamical interactions between hot and cold gas.
More insights to the evolution of early-type galaxies should also come from the
metallicities of all the gas phases.

\section*{Acknowledgements}

\noindent
I acknowledge support from NSF AST 1109803 and travel support from the IAU; I also thank Dr Paul T.~P.~Ho for the
invitation to spend a sabbatical year at ASIAA.

\end{document}